\documentclass[
 preprint,
 superscriptaddress,
 amsmath,amssymb,
 aps,
 prl,
 showpacs,
 floatfix
]{revtex4-1}

\usepackage{graphicx}
\usepackage{dcolumn}
\usepackage{bm}
\usepackage[colorlinks]{hyperref}

\begin{document}


\title{A Cratered Photonic Crystal Cavity Mode for Nonlocal Exciton-Photon Interactions}

\author{Chenjiang Qian}
\author{Xin Xie}
\author{Jingnan Yang}
\affiliation{Beijing National Laboratory for Condensed Matter Physics, Institute of Physics, Chinese Academy of Sciences, Beijing 100190, China}
\affiliation{CAS Center for Excellence in Topological Quantum Computation and School of Physical Sciences, University of Chinese Academy of Sciences, Beijing 100049, China}

\author{Xiulai Xu}
\email{xlxu@iphy.ac.cn}
\affiliation{Beijing National Laboratory for Condensed Matter Physics, Institute of Physics, Chinese Academy of Sciences, Beijing 100190, China}
\affiliation{CAS Center for Excellence in Topological Quantum Computation and School of Physical Sciences, University of Chinese Academy of Sciences, Beijing 100049, China}
\affiliation{Songshan Lake Materials Laboratory, Dongguan, Guangdong 523808, China}


\begin{abstract}

Optical nanocavities for coherent interfaces usually have their electric field maximum at the center point, which normally benefits interactions with small local quantum emitters. Here, we propose a partial thickness modulation on 2D slab photonic crystal cavities for a cratered cavity mode function to improve nonlocal interactions. The thickness modulation is applied around the central region, and has little effect on the fringe electric field, which determines the coupling to waveguides or other cavities. Furthermore, the partial modulation enhances the cratered electric field at positions that are distant from the center point. Therefore, interactions with multiple separated emitters are simultaneously enhanced, and the interaction with a large emitter beyond the dipole approximation is also doubled. The improvement of the nonlocal interactions demonstrates a great potential for the cratered cavity mode profile for applications in quantum photonic networks.

\end{abstract}
\maketitle

\section{\label{sec1} Introduction}

A two dimensional (2D) photonic crystal (PC) from a semiconductor slab with periodic air holes is a mature optical system used for integrating cavities and waveguides on a single chip \cite{Atature2018}, and it provides a good platform for quantum information processing and quantum photonic networks \cite{PhysRevLett.83.4204,Carter2013,Bose2014}. Previous designs of 2D PCs mainly focused on the in-plane XY direction, such as the H0, H1 or L3 cavities, by moving certain air holes \cite{Vahala2003}. The cavity mode function (or electric field) generally contains a periodic function along with an envelope function large in the center and small at the edges. The large central electric field is suitable for interactions with embedded quantum emitters \cite{RevModPhys.87.347}, which serves as the key node in a quantum photonic network \cite{Kimble2008,Ritter2012,RevModPhys.87.1379}. The small fringe electric field determines the coupling to waveguides or other cavities \cite{Chalcraft:11} that connect the multiple nodes.

To improve the exciton-photon interaction, PC cavities have been continuously optimized for a high Q and a small mode volume. The optimizations are mainly through shifts or reducing the size of air holes at the cavity edge \cite{Akahane2003,Song2005,doi:10.1063/1.5016615}. However, these 2D in-plane designs have limitations. Shifting or shrinking air holes at the edge is usually complex and significantly affects the fringe electric field, resulting in side effects on the node connections \cite{Minkov2014,Kuramochi:14,Zhao:15}. More importantly, the small mode volume corresponding to the large maximum electric field only benefits interactions with a small emitter at the antinode (center point) with the dipole approximation (DA) \cite{doi:10.1063/1.4773882}. However, as the quantum photonic network is scaled up, the interactions with multiple emitters become more important \cite{Douglas2015}. Meanwhile, large emitters are widely used for large interaction strengths \cite{Reithmaier2004,PhysRevLett.120.213901}, where the interaction can be beyond the DA \cite{Andersen2010,qianenhanced}. In these nonlocal interactions the electric field at the center point has fewer contributions.

Here, we propose a partial thickness modulation of 2D PC cavities as a 3D design in the Z direction. Different thicknesses can be approximated to different effective refractive indices \cite{doi:10.1063/1.1500774}, resulting in the partial modulation of the cavity mode function. The thickness modulation is only applied around the center region; thus, the central electric field is modulated, while the fringe electric field and node connections are barely affected. Furthermore, the electric field away from the center point is enhanced and even exceeds the value at the center point, forming a cratered cavity mode function. In contrast, this enhancement is hard to achieve by 2D designs. The cratered cavity mode directly benefits interactions with multiple separate emitters. The maximum interaction strength to a large emitter beyond the DA is also doubled. Our work proposes a new approach to modulate 2D PC cavities and demonstrates its necessity for the nonlocal exciton-photon interaction systems. Controlling the slab thickness is feasible after fabricating the PC; thus our work has great potential for the building and optimization of quantum photonic networks.

\section{\label{sec2} Thickness Modulation}

\begin{figure}
\includegraphics[scale=0.7]{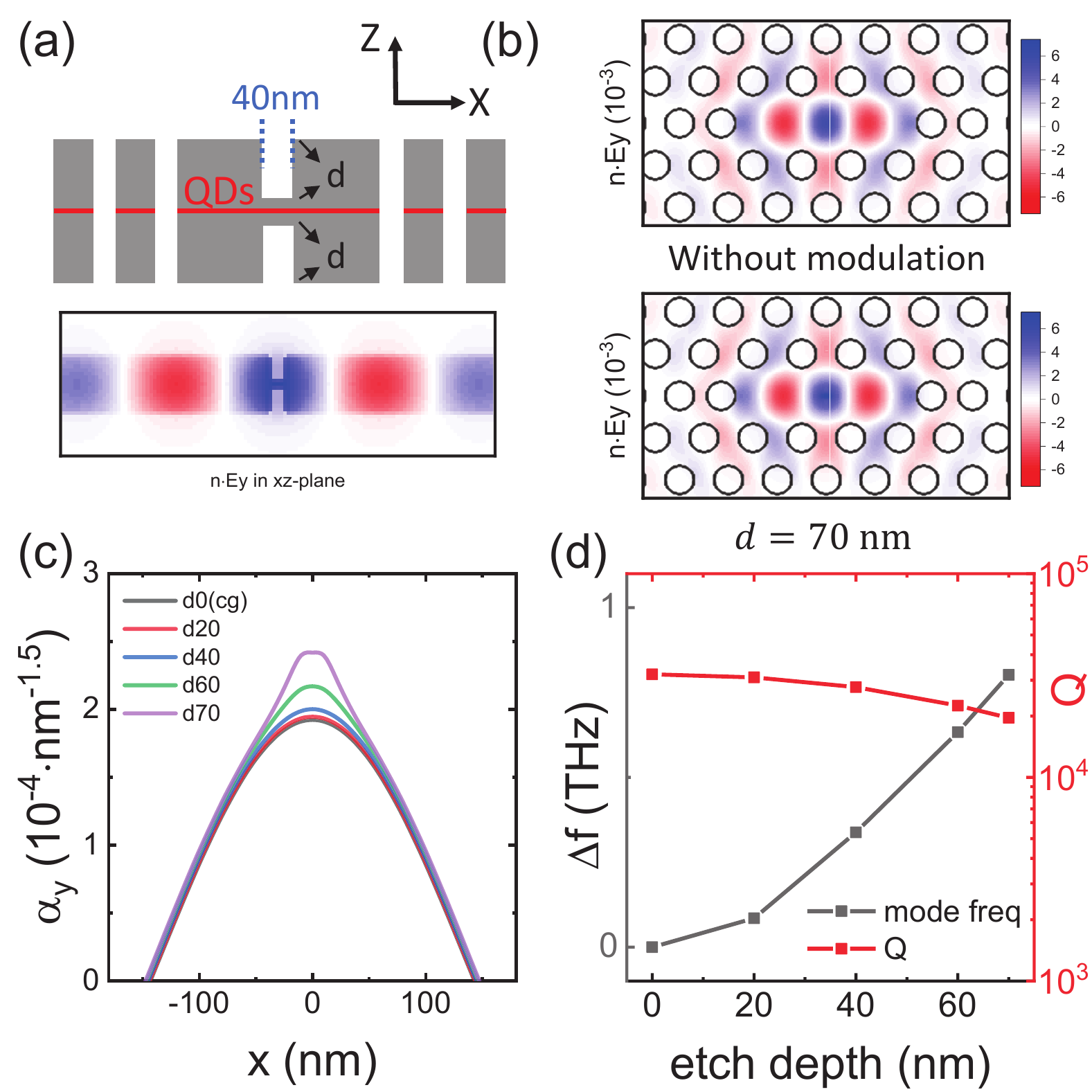}
\caption{\label{f1} (a) (Top) Schematic of the L3 cavity with partial thickness modulation. (Bottom) The electric field distribution $nE_y$ in the XZ plane. (b) The electric field distribution $nE_y$ at $z=0$ without modulation $d=0\ \mathrm{nm}$ (top)  and with  $d=70\ \mathrm{nm}$ (bottom). (c) The cavity mode function $\alpha_y$ at $y,z=0$. (d) Mode frequency and Q as a function of the depth $d$.}
\end{figure}

The thickness modulation in our work is implemented on an L3 cavity on a GaAs slab with a refractive index of $3.45$.  The GaAs slab is set in the XY plane, and the origin of the coordinates is set at the cavities center point. The lattice constant of the photonic crystal structure $a$ is $320\ \mathrm{nm}$ and the slab thickness is $160\ \mathrm{nm}$. The radius of the air holes is $0.29\ a$. The two air holes at the cavities edge are shifted by $0.15\ a$. This structure is similar to the first high-Q PC cavity that was achieved \cite{Akahane2003}. Contrary to most further optimizations with shifts and size adjustments of more air holes in-plane \cite{Minkov2014,Kuramochi:14}, we partially modulate the slab thickness as shown in Fig.~\ref{f1}(a), which can be realized by an additional etching in the experiment. The smaller thickness can be described with a smaller refractive index in the effective refractive index approximation for 2D slab optical structures \cite{doi:10.1063/1.1500774,Amann:81}. Similar to a larger electron wave-function with a smaller electric potential in the crystal, the partial enhancement of the cavity mode function $\alpha$ can be predicted from the smaller effective refractive index in the photonic crystal. The detailed cavity mode function of the first high-Q mode is then calculated by the 3D finite-different time-domain (FDTD) method.

The thickness modulation is first applied in a circular region with a radius of $20\ \mathrm{nm}$ from the center point. The etching depth $d$ is set to different values. Figure~\ref{f1}(a) shows the structural schematic and the normalized electric field $nE_y$ in the XZ plane with $d=70\ \mathrm{nm}$. $n$ is the refractive index with a value of 3.45 for GaAs and 1 for air. The electric field is compressed in the Z direction with an enhancement at $z=0$ where the emitters are usually embedded. A comparison of $nE_y$ in the XY plane at $z=0$ is shown in Fig.~\ref{f1}(b), revealing the modulation only around the center region and not at the cavities edge. Meanwhile, $nE_y$ is symmetric at both the X and Y directions. Thus, the cavity mode polarization is along the Y direction, which is a typical feature of an L3 cavity and is not affected by the modulation \cite{doi:10.1063/1.2748310}. Figure~\ref{f1}(c) shows the cavity mode function $\alpha_y$ with different values of $d$. The result of the original cavity without modulation is labeled as the control group (cg). The original frequency $f_{cg}$ is $256.493\ \mathrm{THz}$, and $\alpha_{y,cg}$ at the center point is $1.9\times 10^{-4}\ \mathrm{nm}^{-1.5}$, corresponding well to the $|\alpha |_{max}=1/\sqrt{V}$ with a mode volume $V\approx(\lambda/n)^3$ for the L3 cavity. With $d=70\ \mathrm{nm}$, a $27\%$ enhancement of $\alpha_y$ at the center point is obtained. This enhancement leads to an equal enhancement of the interaction strength to a small emitter at the center point with the DA \cite{book1}.

Additionally, the cavity mode frequency and Q were changed by the modulation, as shown in Fig.~\ref{f1}(d). The frequency shift is approximately $1\ \mathrm{THz}$, which is quite small compared to $f_{cg}$. This means that the additional detuning caused by the modulation can be easily cancelled out by controls on the cavity or emitter, such as the gas condensation, temperature, electric field and magnetic field \cite{PhysRevB.65.041308,doi:10.1063/1.2076435,O'Brien2009,PhysRevLett.104.047402}. Meanwhile, if the emitter emission frequency is already known, then the additional detuning can also be compensated by adjusting the lattice constant $a$. The cavity Q decays from 30000 to 20000. This decay is also relatively small and acceptable, especially compared to another additional decay where the cavity is in a realistic PC network. The node connections in a PC network usually result in the decay of Q by an order of magnitude \cite{Kuramochi2014,photonics3040055}. Therefore, by the partial modulation the exciton-photon interactions can be enhanced with little side effect on the cavity mode and node connections in the network.

\section{\label{sec3} Cratered Enhancement}

\begin{figure}
\includegraphics[scale=0.7]{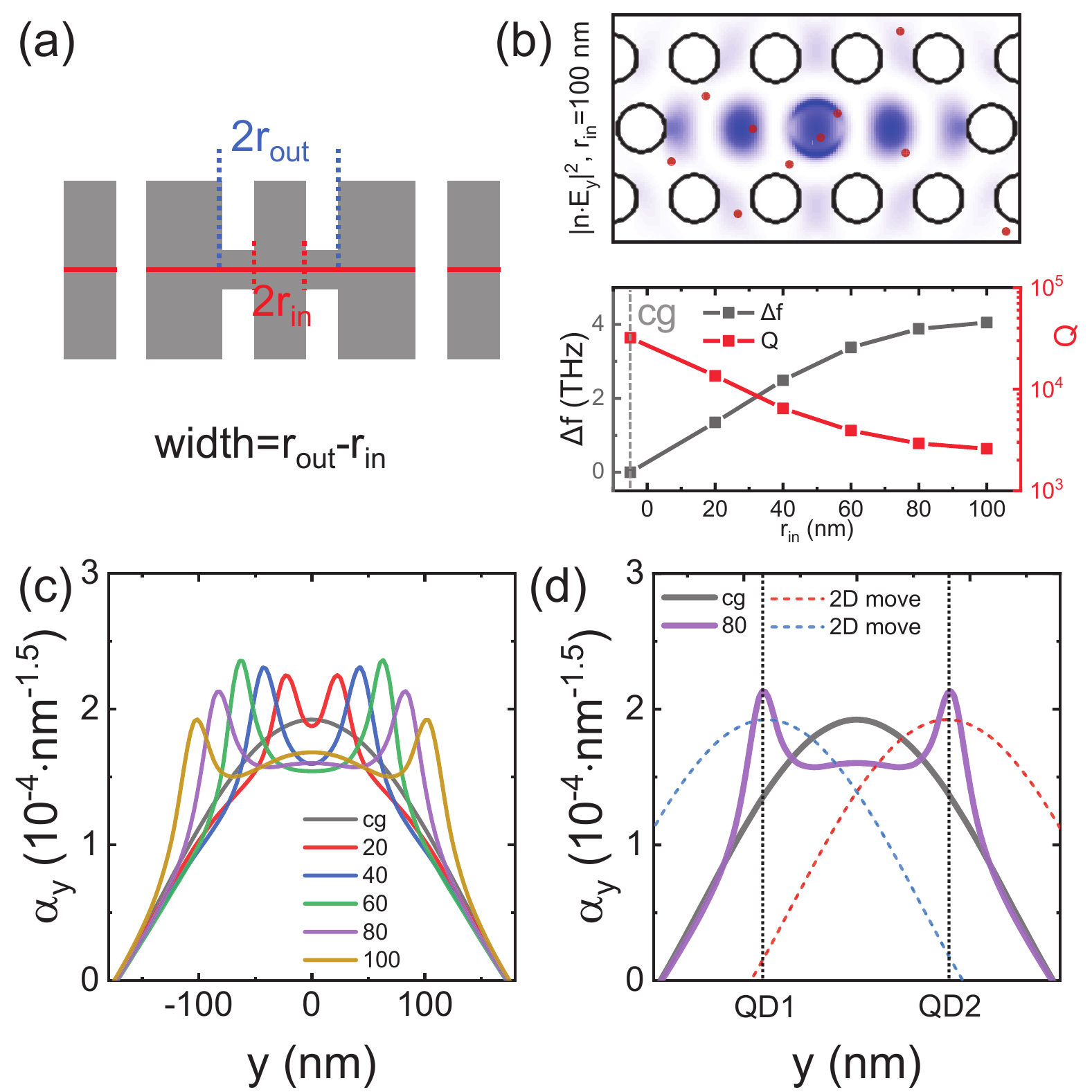}
\caption{\label{f2} (a) Schematic of the ring etching with an $r_{in}$ and $r_{out}$. (b) (Top) The energy distribution $|nE_y|^2$ at $z=0$ with $r_{in}$ of 60 and 100 $\mathrm{nm}$. The red dots refer to randomly positioned emitters. (Bottom) Mode frequency and Q with different values of $r_{in}$. (c) The cavity mode function $\alpha_y$ at $x,z=0$. (d) A comparison between the 3D and 2D designs.}
\end{figure}

The electric field enhancement in Fig.~\ref{f1} can improve interactions with a small emitter at the center point. In reality, emitters such as quantum dots (QDs) are usually randomly positioned and separated \cite{book2}. Additionally, the exciton-photon interactions with a large emitter can be beyond the DA \cite{qianenhanced}. In these nonlocal interactions the electric field enhancement away from, rather than at, the center point is of great importance. Therefore, as shown in Fig.~\ref{f2}(a), the circle etching evolves to a ring etching with an etching depth of $70\ \mathrm{nm}$, a ring width of $10\ \mathrm{nm}$ and a different inner radius $r_{in}$ that ranges from 20 to 100 $\mathrm{nm}$. Figure~\ref{f2}(b) shows the mode frequency and Q with different values of $r_{in}$, along with the electric field $nE_y$ with $r_{in}=100\ \mathrm{nm}$. The enhancement of $nE_y$ corresponding to the ring etching can be clearly observed. The cavity mode function $\alpha_y$ is then cratered as also shown in Fig.~\ref{f2}(c). $\alpha_y$ away from the center point is enhanced with values exceeding that at the center point. This is a great advantage compared to normal 2D designs, where $|\alpha|_{max}$ is always at the antinodes with a period of a half wavelength. The cratered cavity mode function greatly benefits the interactions with multiple separated emitters. For example, as shown in Fig.~\ref{f2}(d), two QDs with positions at $y=80\ \mathrm{nm}$ and $y=-80\ \mathrm{nm}$ and a distance of approximately a quarter wavelength can be simultaneously enhanced by an electric field with $r_{in}=80\ \mathrm{nm}$ (purple line). By moving the cavity center with 2D designs (dashed lines), if one emitter is at the antinode with the maximum interaction strength, interactions with the other emitter would decay to zero. These results demonstrate the advantage of the cratered cavity mode for interactions with multiple separated emitters, which is significant for scaling up quantum photonic networks. In contrast, the limitation of a normal cavity mode without modulation is obviously more significant when the number of emitters continuously increases.

In addition to the interactions with multiple emitters discussed above, interactions beyond the DA with a large emitter are also strongly related to the mode function away from the center point. Large emitters are usually chosen for their large dipole moment to improve the exciton-photon interaction \cite{Reithmaier2004}. The interaction strength generally cannot infinitely increase with emitter size and has a maximum value due to the breakdown of the DA. Previous work gives a strict calculation of the interaction strength with the unit cell dipole approximation instead of the DA in the entire emitter \cite{PhysRevB.86.085304}. For brevity we set complicated constant parameters to 1, and then we get the interaction strength $ g\propto|\int d^3\textbf{r}\chi^\ast(\textbf{r}_0,\textbf{r},\textbf{r}) \alpha_y(\textbf{r})|$, where $\chi^\ast(\textbf{r}_0,\textbf{r},\textbf{r})$ is the exciton envelope wave-function with the emitter center at $\textbf{r}_0$ and an electron hole pair at $\textbf{r}$. $\alpha_y(\textbf{r})$ can be extracted from the calculation results, while $\chi^\ast(\textbf{r}_0,\textbf{r},\textbf{r})$ is quite complex. For example, the wave-function simulation of QDs is always nontrivial and has various theoretical models with different results \cite{PhysRevB.54.R2300,PhysRevB.59.5688,PhysRevLett.91.056404,PhysRevLett.96.187602}.

\begin{figure}
\includegraphics[scale=0.7]{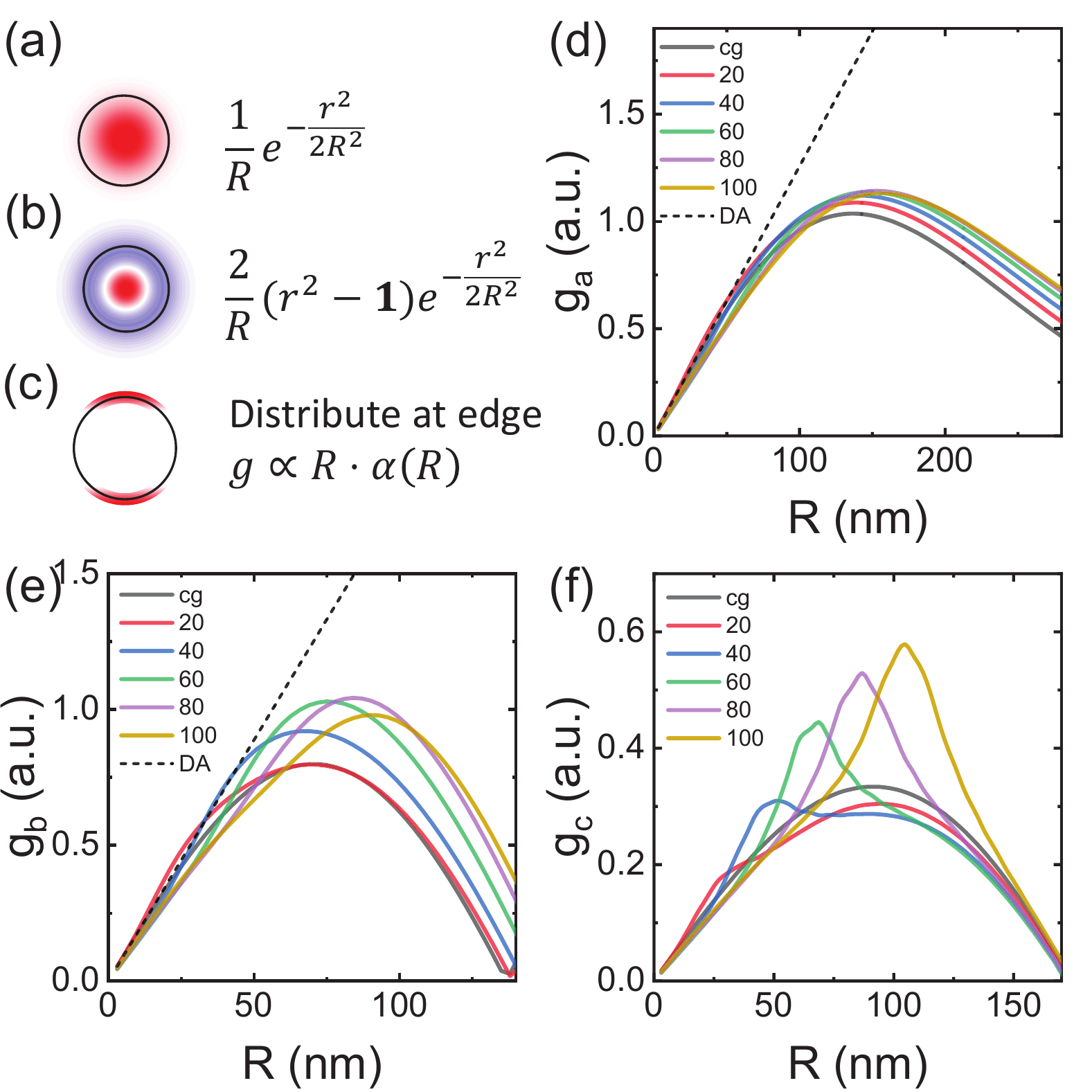}
\caption{\label{f3} (a)-(c) Three types of QD wave-functions with their schematics. (d)-(f) The size-dependent interaction strength for QD with the first (d), second (e) and third (f) types of wave-function.}
\end{figure}

The interaction strength is calculated with three types of wave-functions of a quantum emitter, as shown in Fig.~\ref{f3}(a)-(c). The emitter is set at the center $\textbf{r}_0=0$, and it has a different radius of size $R$. For large disk-shaped QDs in a weak confinement regime, the wave-function in the XY plane is Hermite polynomials as the solution of an isotropic harmonic oscillator \cite{PhysRevB.45.11036}. We take the first two even-parity Hermite polynomials as the ground (Fig.~\ref{f3}(a)) and the excited (Fig.~\ref{f3}(b)) states. These two wave-functions have centralized distributions, which means $|\chi|_{max}$ is at the QD center. This is different from realistic QDs \cite{Vdovin122,RevModPhys.76.725,PhysRevLett.108.046801}. In contrast, another limiting case is that all wave-functions distributed around the edge are similar to quantum rings. In this case, the interaction strength can be easily calculated with a modified dipole approximation. The dipole moment is proportional to $R$ and the mode function is taken with the value at the QD edge $\alpha_y(R)$ instead of $\alpha_y(0)$, as shown in Fig.~\ref{f3}(c). Figure~\ref{f3}(d)(e) show the calculated interaction strengths with the ground ($g_a$) and excited ($g_b$) states in the harmonic oscillator model. The maximum $g_a$ is approximately $R=150\ \mathrm{nm}$, which is similar to previous calculation results \cite{PhysRevB.86.085304}. The maximum $g_b$ is approximately $R=75\ \mathrm{nm}$. This demonstrates that the nonlocal effect of excited states is much more significant than that of the ground states, which corresponds well with recent experimental results \cite{qianenhanced}. The cratered cavity mode has several effects on $g_a$, while the maximum of $g_b$ has a $31\%$ increase with an $r_{in}$ of $80\ \mathrm{nm}$. This is because $\chi$ in the excited state is larger than in the ground state. Similarly, in the third case the interaction strength $g_c$ of $\chi$ with a much larger extent has a much larger (doubled) enhancement, as shown in Fig.~\ref{f3}(f). These results demonstrate the cratered cavity mode with great improvements to the exciton-photon interactions beyond the DA. Nonetheless, it should be noted that all three types of wave-functions are simple ideal models, which may be quite different from realistic emitters. The theoretical calculations mainly give a qualitative result, while the quantitative characteristics in realistic cavity-dot systems need an experimental demonstration.

\section{\label{sec4} Optimization and Feasibility}

\begin{figure}
\includegraphics[scale=0.7]{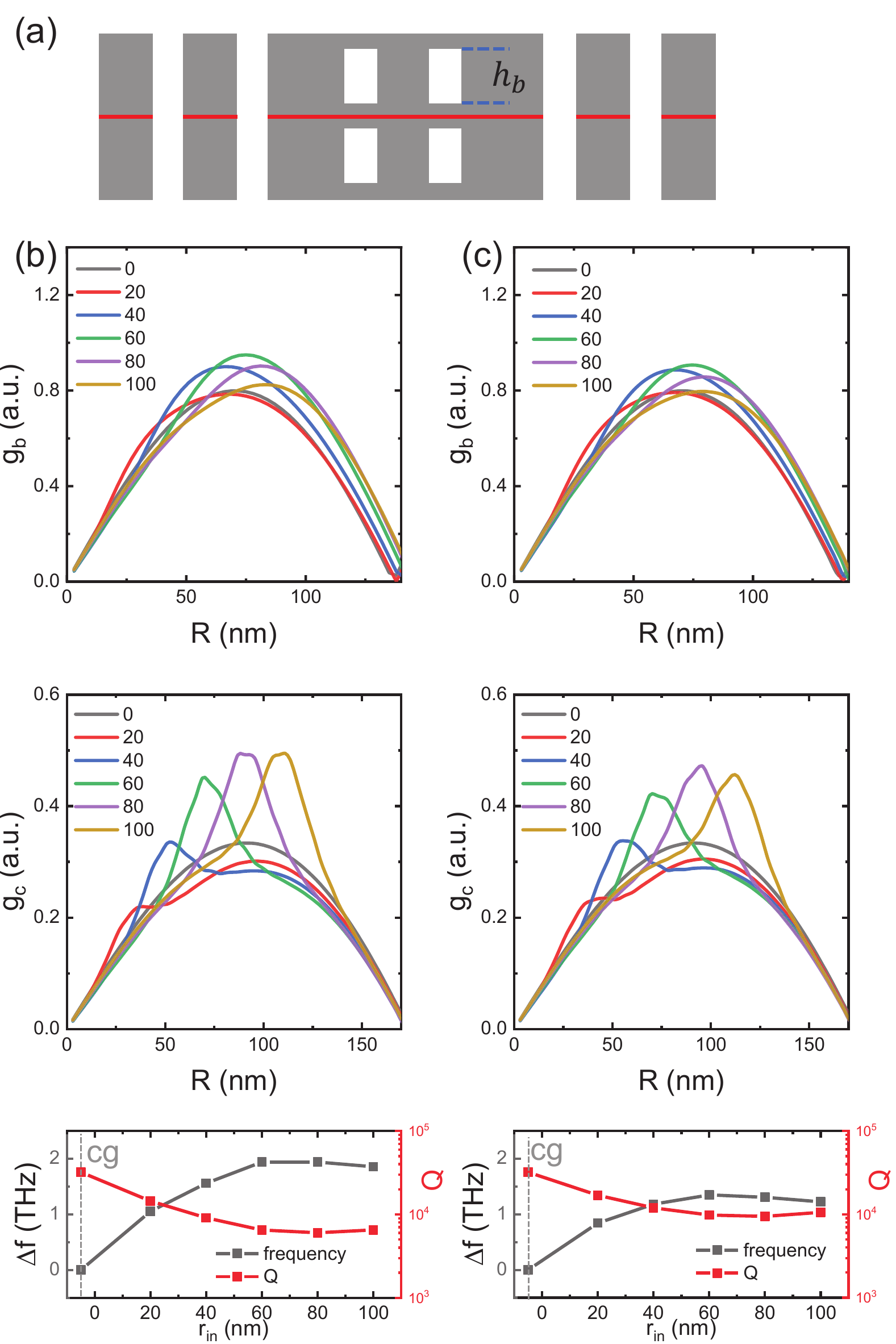}
\caption{\label{f4} (a) Schematic of the optimization with control of $h_b$. (b)(c) The calculation results of the interaction strength beyond the DA, mode frequency and cavity Q with (b) $h_b$ of $30\ \mathrm{nm}$ and ring width of $20\ \mathrm{nm}$, and (c) $h_b$ of $20\ \mathrm{nm}$ and ring width of $25\ \mathrm{nm}$. The cavity Q is greatly improved.}
\end{figure}

The ring etching in Fig.~\ref{f2} results in a decay of Q due to the large etching area with a large $r_{in}$. The decay is approximately an order of magnitude, similar to that originating from the node connections in the network \cite{Kuramochi2014,photonics3040055}. Thus, the modulation would be more perfect with further optimizations to improve the cavity Q. For interaction with multiple small emitters the optimization is easy. Only the electric field at the emitter positions, instead of the entire ring, needs to be enhanced. Thus, the cavity Q can be significantly increased with a decreased etching area. For the exciton-photon interactions beyond the DA the electric field enhancement in the entire ring is important. The optimization is then performed with the height of the ring etching $h_b$, as shown in Fig.~\ref{f4}(a). Such structures can be realized by direct laser writing \cite{ZHANG2010435} or by an additional multilayer growth of material in the ring etching region. Fig.~\ref{f4}(b) shows the calculation results with an $h_b$ of $30\ \mathrm{nm}$, a ring width of $20\ \mathrm{nm}$ and different values of $r_{in}$. Compared to the results in Fig.~\ref{f3}, the enhancement of $g_b$ is smaller, while the enhancement of $g_c$ is still significant. More importantly, the cavity Q is improved with values over 6000, which is large enough for the strong interaction \cite{Yoshie2004,doi:10.1063/1.3487937}. Fig.~\ref{f4}(c) shows the calculation results with an $h_b$ of $20\ \mathrm{nm}$ and a ring width of $25\ \mathrm{nm}$. In this case, the cavity Q is further improved with values over 10000, with a smaller interaction enhancement. Therefore, the control of $h_b$ results in an acceptable cavity Q and a still significant enhancement of exciton-photon interactions beyond the DA.

\begin{figure}
\includegraphics[scale=0.7]{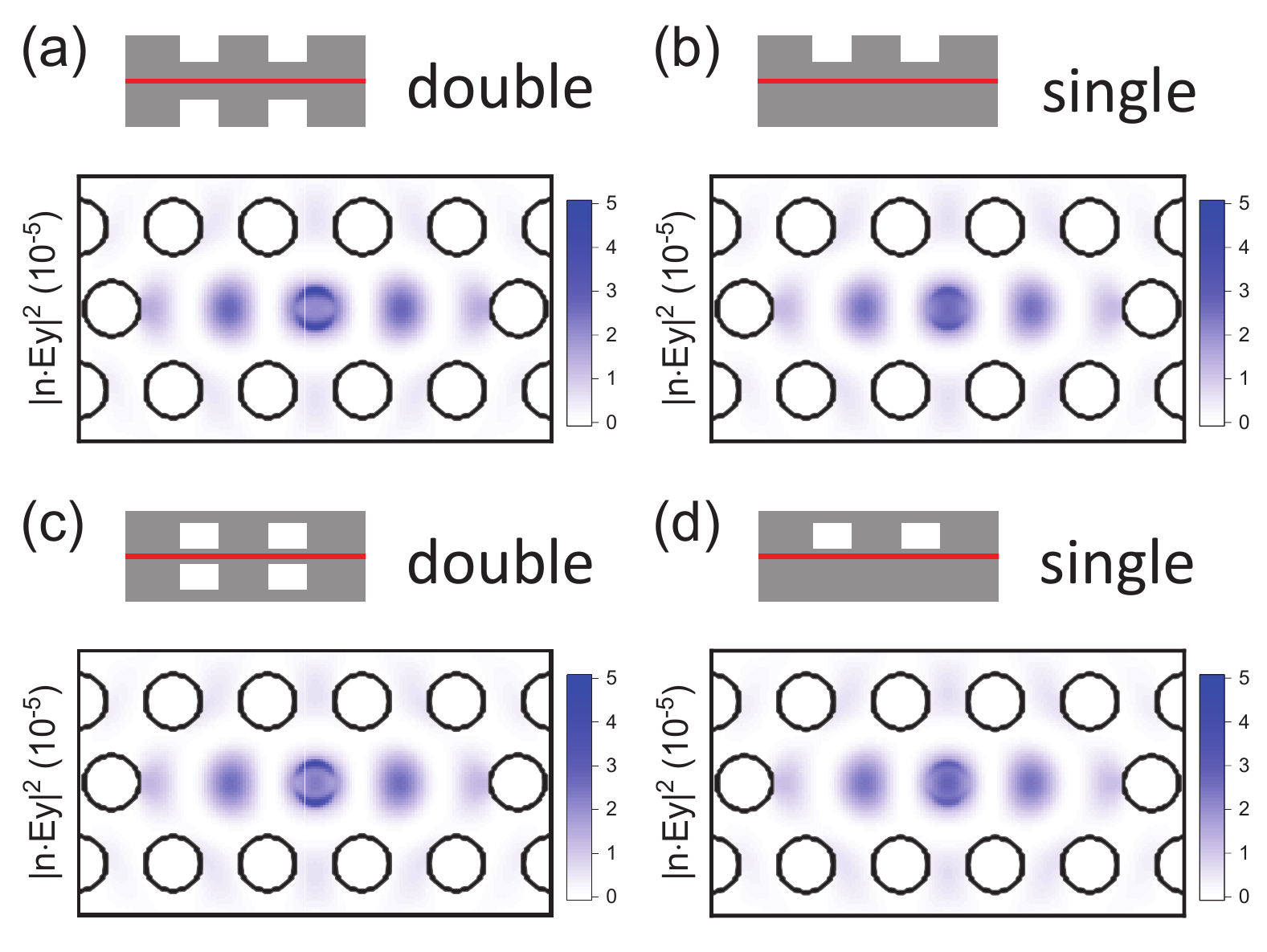}
\caption{\label{f5} (a)(b) Schematic and energy distribution $|nE_y|^2$ with the etching bar on (a) both sides and (b) a single side, corresponding to the results in Fig.~\ref{f2}(b) with $r_{in}=60\ \mathrm{nm}$. (c)(d) The schematic and energy distribution $|nE_y|^2$ with the optimized etching bar on (c) both sides and (d) a single side, corresponding to the results in Fig.~\ref{f4}(b) with $r_{in}=60\ \mathrm{nm}$.}
\end{figure}

The calculations above are performed with etching bars on both sides in the Z direction, as shown in Fig.~\ref{f5}(a)(c). This can be achieved by a predesign of the sample structure or another complex fabrication. The thin central section might also lead to mechanical instability. These problems can be solved by optimizing for single side etching, as shown in Fig.~\ref{f5}(b)(d). Single side etching can be easily applied with a second lithography in a two-patterning process \cite{doi:10.1116/1.3308972}. Such an approach applied after fabricating the PC is especially suitable for emitters with a random position. Meanwhile, the single side etching results in a thicker central section and thus improves the mechanical stability. Figure~\ref{f5}(b)(d) shows the simulation results of the single side etching. A similar cratered cavity mode at the ring etching is observed, indicating similar effects on the exciton-photon interactions as the double side etching. The fringe electric field at the cavity edge is also barely affected. Thus, by utilizing the single side etching, the feasibility of the thickness control is greatly improved, along with similar improvements to the nonlocal interactions. It is worth noting that these optimizations are a theoretical proposal. In real experiments the fabrication process and mechanical stability should depend on the materials. Thus, a more specific optimization for specific materials should be performed in the future. Nevertheless, the design and the proposal should work for a system to improve the interaction strength.

\section{\label{sec5}Conclusion}

In summary, we demonstrated a cratered cavity mode function to improve nonlocal interactions with multiple separated emitters or a large emitter beyond the DA. In the nonlocal interactions the optimization of the cavity mode distribution is more significant than the normal optimization for a high Q or small mode volume. We proposed a thickness modulation of 2D PC cavities and achieved the cratered cavity mode with a doubled nonlocal interaction strength. Meanwhile, the fringe electric field at the cavity edge is barely affected, resulting in hardly any side effects to the node connections in the network. Additionally, the thickness modulation is highly feasible with current fabrication technology. These results indicate the advantages of the new 3D design compared to normal in-plane optimizations. As the nonlocal interactions become more important in quantum information processing, we believe our work can benefit cavity quantum electrodynamics with a great potential for building a quantum photonic network.

\section{Acknowledgments}
This work was supported by the National Natural Science Foundation of China under Grants No. 61675228, No. 11721404, No. 51761145104 and No. 11874419; the Strategic Priority Research Program, the Instrument Developing Project and the Interdisciplinary Innovation Team of the Chinese Academy of Sciences under Grants No. XDB28000000 and No.YJKYYQ20180036.

\section{Conflict of Interest}
The authors declare no conflict of interest.

\section{Keywords}
Cavity quantum electrodynamics, Photonic crystal cavity, Nonlocal interactions

%

\end{document}